\documentclass[prb,showpacs,preprintnumbers,amsmath,amssymb]{revtex4}

\usepackage{graphicx}
\usepackage[space]{grffile}
\usepackage{latexsym}
\usepackage{amsfonts,amsmath,amssymb}
\usepackage{url}
\usepackage[utf8]{inputenc}
\usepackage{fancyref}
\usepackage{hyperref}
\hypersetup{colorlinks=false,pdfborder={0 0 0},}
\usepackage{textcomp}
\usepackage{longtable}
\usepackage{multirow,booktabs}
\usepackage{natbib}

\begin{document}

\title{Simulating nanoscale heat transport}

\author{Giuseppe Romano}
\email{romanog@mit.edu}
\affiliation{Department of Materials Science, Massachusetts Institute of Technology, 77 Massachusetts Avenue, Cambridge (MA), 02139}
\author{Jean-Philippe Peraud}
\email{jperaud@mit.edu}
\affiliation{Department of Mechanical Engineering, Massachusetts Institute of Technology, 77 Massachusetts Avenue, Cambridge (MA), 02139}
\author{Jeffrey C. Grossman}
\email{jcg@mit.edu}
\affiliation{Department of Materials Science, Massachusetts Institute of Technology, 77 Massachusetts Avenue, Cambridge (MA), 02139}
\maketitle




\section{Introduction}
Heat conduction has been modeled for almost two centuries by the well known Fourier's law. Jean-Baptiste Joseph Fourier stated that "the quantity of heat which flows uniformly, during unit of time, across unit of surface taken on any section whatever parallel to the sides, all other things being equal, is directly proportional to the difference of the extreme temperatures, and inversely proportional to the distance which separates these sides"~\cite{Fourier_2009}. Fourier's law can be conveniently written in its local form
\begin{equation}
\mathbf{J} = -\kappa \nabla T, \label{fourier}
\end{equation}
where $\kappa$ is the thermal conductivity (for Silicon it is about $150 Wm^{-1}K^{-1} $), $T$ is the lattice temperature and $\mathbf{J}$ is the heat flux. Eq.~\ref{fourier} together with the continuity equation for thermal flux 
\begin{equation}
\rho C_p \frac{\partial T}{\partial t} + \nabla \cdot \mathbf{J} = H, \label{conduction}
\end{equation}
gives the well known diffusive heat conduction equation. In Eq.~\ref{conduction}, $C_p$ is the specific heat capacity, $\rho$ the material density and $H$ the sum of all the heat sources in the system.

While Fourier's law is valid at the macroscale, it has been experimentally proved that in some nanostructured materials, such as nanowires~\cite{Li_2003}, thin films~\cite{Ju_1999} or nanoporous materials~\cite{Song_2004}, the description of thermal transport as diffusive heat conduction breaks down. Furthermore, as we will see later, Fourier's law is violated in transient heat conduction experiments~\cite{Maasilta_2014}\cite{Wilson_2014}.

The departure of heat transport from the diffusive regime can be qualitatively explained by considering heat as being carried by material waves, with their own frequency, group velocity and dispersion curve~\cite{chen2005nanoscale}. Wave effects can be taken into account by the so-called molecular dynamics (MD) simulations, which essentially apply Newton's Law to a system of atoms, whose internal forces are modeled by either empirical or first-principles calculations. The thermal conductivity computed by a MD-based formalism inherently includes both harmonic and anharmonic effects and can be computed by the the well-known Green-Kubo formula
\begin{equation}\label{moldynflux}
\kappa_{\alpha \beta} = \frac{1}{V k_B T^2}\int_0^\infty <J_\alpha (0) J_{\beta}(t) >  dt
\end{equation}
where $k_B$ is the Boltzmann constant, V the volume of the system and $\kappa_{\alpha \beta}$ is the thermal conductivity tensor. In Eq.~\ref{moldynflux}, $<A>$ is the average ensemble of the observable $A$, and can be replaced by a time average if the simulation is long enough to satisfy ergodicity~\cite{Esfarjani_2011}. Although MD is a powerful approach to model thermal conductivity at very small scales, it can become computationally prohibitive when dealing with realistic structures of mesoscale dimensions, such as thin films and porous materials. To overcome this limit, one may model the quantized lattice vibrations - or phonons - as particles, and compute the temperature as well as the thermal fluxes by applying numerical schemes solving for particle transport problems. 

 This simplification is justified by noting that at room temperature the dominant phonons in certain materials, such as Si, have wavelengths below $10~nm$. Within this assumption, phonon transport can be safely modeled by the Boltzmann Transport Equation (BTE)~\cite{Ziman_2001}. The BTE has been gaining much attention recently in modeling thermal transport in nanostructured materials, especially for thermoelectric applications. In the following, we will focus on the BTE and the two main approaches used to solve it in complex structures. Readers interested in modeling wave effects at the mesoscale can refer to~\cite{Davis_2014}. Also, while here we discuss the potential of the BTE in nanoscale heat transport simulations, there are many equally important methods left aside. A comprehensive review on computational tools for heat transfer can be found in~\cite{Chen_2014}.

\section{The Boltzmann Transport Equation}
In order to introduce the BTE for phonons, we first define $f(\mathbf{r},\mathbf{k},t)$ as the phonon distribution function at a given time $t$, at position $\mathbf{r}$ with wave vector $\mathbf{k}$. With no loss of generality, we assume the Brillouen zone to be isotropic. We further define the quantity $I(\mathbf{r},\mathbf{s},\omega,p)=\frac{1}{4\pi}|\mathbf{v(\omega,p)}| f D(\omega,p) \hbar \omega$, where $D(\omega,p)$ is the density of states, $\mathbf{v}(\omega,p)$ is the group velocity, $\hbar \omega$ the phonon energy and $p$ the phonon branch. The group velocity can be obtained from the dispersion relations $\omega(\mathbf{k})$ as $\mathbf{v}=\frac{\partial \omega(\mathbf{k})}{\partial \mathbf{k}}$. The quantity $I(\mathbf{r},\mathbf{s},\omega,p)$ represents the intensity of phonons for a given frequency within a unit solid angle~\cite{Majumdar_1993}, traveling along the direction $\mathbf{s} = \frac{\mathbf{v}}{|\mathbf{v}|}$. In absence of any external driving forces, such as an applied temperature gradient, the intensity $I(\mathbf{r},\mathbf{s},\omega,p)$ equals the equilibrium intensity $I_0(T_0)$, defined as $I_0(T_0)=\frac{1}{4\pi}|\mathbf{v(\omega,p)}| f_0(\omega,T_0) D(\omega,p) \hbar \omega$. The term $f_0(T_0)$ is the Bose-Einstein distribution $f_0(T_0) = [\exp(\frac{\hbar \omega}{k_B T_0})-1]^{-1}$. Assuming that all phonons scattering events are uncorrelated, the non-equilibrium phonon intensity can be modeled by
\begin{equation}\label{bte}
\frac{1}{|\mathbf{v}|}\frac{\partial I}{\partial t} + \mathbf{s}\cdot \nabla I = \frac{I_0(T)-I}{\tau |\mathbf{v}|},
\end{equation}
which is the BTE under the so-called relaxation time approximation~\cite{Majumdar_1993}. In Eq.~\ref{bte}, $\tau$ is the intrinsic scattering time~\cite{broido2007intrinsic}. Energy conservation among modes can be obtained by applying the continuity equation for the energy flux, which results in
\begin{equation}\label{econs}
\sum_{p}  \int_0^{\omega_M^p} \frac{I_0(T)}{\tau} d\omega = \sum_p \int_0^{\omega_M^p} \frac{ < I >  }{\tau}d\omega,
\end{equation} 
where $I_0(T)$ is the Bose-Einstein distribution at the temperature $T$, $\omega$ is the phonon angular frequency, $p$ is the phonon branch and $\omega_M^p$ is the maximum angular frequency for the branch $p$. In practical thermal conductivity calculations, a difference of temperature $\Delta T$ is applied to a simulation domain and, once Eqs.~\ref{bte}-\ref{econs} are solved consistently, the thermal flux is computed by
\begin{equation}
\mathbf{J}(\mathbf{r})= 4 \pi \sum_p  \int_0^{\omega_M^p}|\mathbf{v}| <I\mathbf{s}> d\omega, 
\end{equation}\label{eflux}
and the {\it effective} thermal conductivity is deduced by $J=-\kappa_{eff} \nabla T$. The BTE described in Eq.~\ref{bte} has to be solved for the whole phonon spectrum and is called the \textit{frequency-dependent} BTE. Generally speaking, the BTE can be solved either deterministically or stochastically. In the following, we devote a section to each approach.

\section{Deterministic solution of the BTE}

Here we show how the BTE can be useful for calculating the steady state thermal conductivity values in nanostructured materials. Following the approach described in~\cite{romano2013multiscale}, we consider only steady state transport. Furthermore, we assume that the bulk thermal conductivity is isotropic and that a very small applied temperature is applied across the sample. Under these simplifications, the BTE becomes
\begin{equation}\label{mfpbte}
\Lambda \mathbf{s} \cdot \nabla   \tilde{T} + \tilde{T} = \gamma \int_0^\infty   \frac{K}{\Lambda'^2}   <\tilde{T}>  d\Lambda',
\end{equation}
where $\tilde{T}$ is the departure of a temperature associated with a given phonon mode from the equilibrium, normalized by the applied temperature. In Eq.~\ref{mfpbte}, the term $K(\Lambda)$ is the bulk phonon mean free path distribution, a quantity that can be obtained either theoretically~\cite{Esfarjani_2011} or experimentally~\cite{minnich2011thermal}-\cite{Regner_2013}, and $\gamma = \left[\sum_p \int_0^\infty  \frac{K}{\Lambda^2} d\Lambda \right]^{-1}$ is a material property, which for Si is $\gamma_{Si} = 2.2739\cdot 10^{-17} m^3W^{-1}K$. The notation $<x>$ stands for the angular average $<x>=\frac{1}{4\pi}\int_{4\pi}x d\Omega$. The right hand side of Eq.~\ref{mfpbte} is related to the normalized {\it effective} lattice temperature.  

Practically, Eq.~\ref{mfpbte} requires the discretization of $K(\Lambda)$, as opposed to the discretization of the phonon frequencies typically used in a frequency-dependent approach. For each MFP, the BTE is solved by means of the Finite Volume method whereas the solid angle is discretized by the Discrete Ordinate Method~\cite{chandrasekhar1960radiative}. Eq.~\ref{mfpbte} is named phonon MFP-BTE and has the following advantages:
\begin{itemize}
\item[-]	It retains the accuracy of the frequency-dependent approach.
\item[-]	It requires the knowledge of only the bulk MFP distribution, which is a quantity that can be obtained from experiments~\cite{minnich2011thermal}. 
\item[-]	The requirements in the discretization of the bulk MFP distribution are less demanding than in a typical frequency-dependent approach, leading to a significant improvement in the computational efficiency~\cite{romano2013multiscale}.
\item[-]  It is relatively easy to parallelize. In fact, each phonon mode is only coupled through the integral appearing on the right hand side of Eq.~\ref{mfpbte}. In a typical iterative solver, each BTE for a given MFP can be run independently, using the effective lattice temperature from the previous step. 

\end{itemize}
In the following, we show an application of the MFP-BTE to nanoporous Silicon, a promising material for thermoelectric applications, thanks to its capability to suppress thermal transport with little degradation in the electrical conductivity~\cite{Song_2004}.

\begin{figure}[h!]
\begin{center}
\includegraphics[width=0.42\columnwidth]{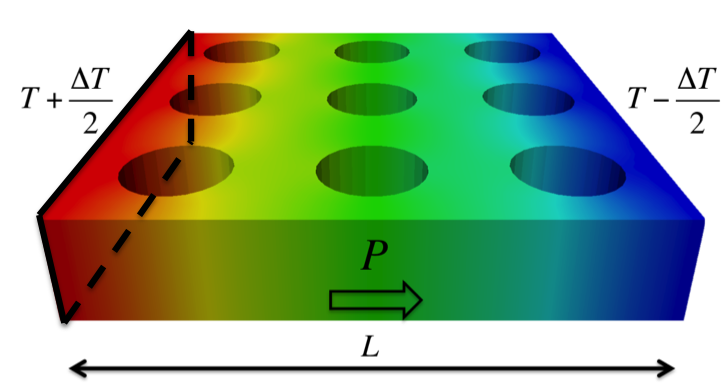}
\caption{\label{domain}
Simulation of a porous material. A difference of temperature $\Delta T$ is applied across the two ends of the domain. }
\end{center}
\end{figure}

\begin{figure}[h!]
\begin{center}
\includegraphics[width=0.84\columnwidth]{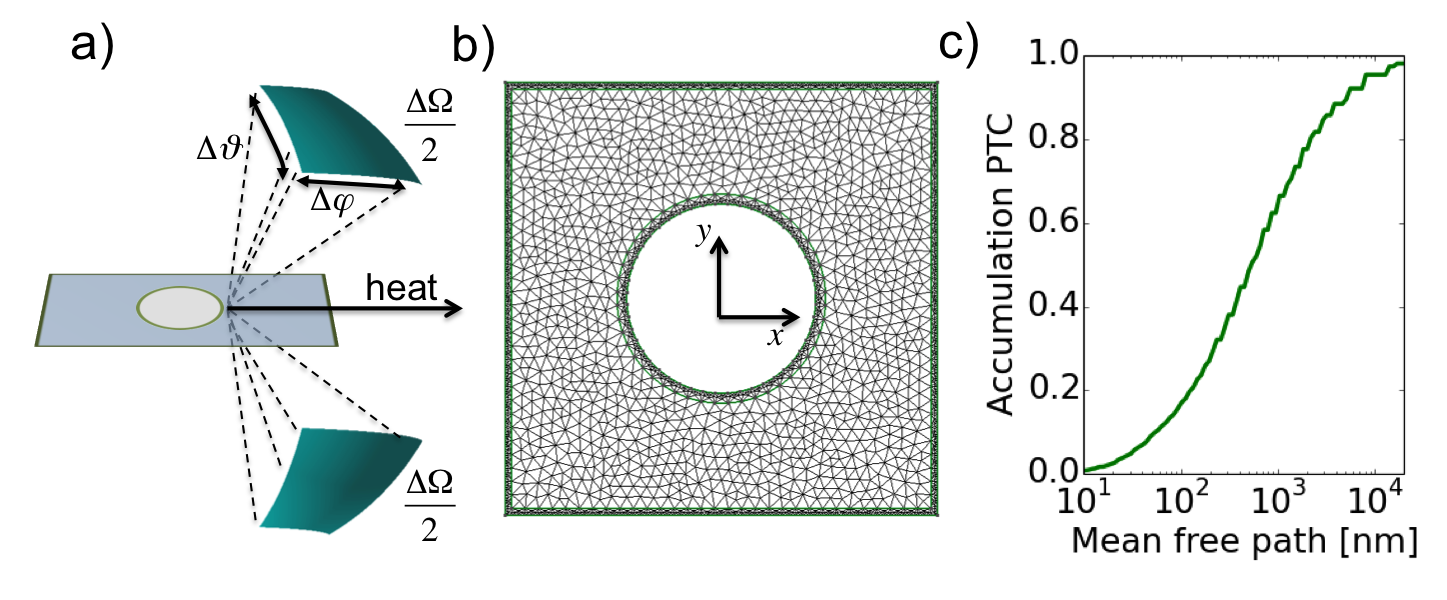}
\caption{\label{discr} a) Uniform scheme for solid angle discretization b) Discretization of the spatial domain c) Cumulative thermal conductivity for Si obtained by a frequency-dependent model. We acknowledge ASME for granting us the permission for using Fig. 1 of the article ``G. Romano and J. C. Grossman. \textit{Journal of Heat Transfer} 137.7 (2015): 071302.''}
\end{center}
\end{figure}

As shown in Fig.~\ref{domain}, a difference of temperature $\Delta T$ is applied to the simulation domain. Then, after the MFP-BTE converges, the thermal power, P, is computed on either the cold or hot side. The effective thermal conductivity $\kappa_{eff}$ is computed by using Fourier's law $\kappa_{eff} = \frac{PL}{A\Delta T}$, where $A$ is the contact area.
In Fig.~\ref{discr}-a the discretization of the solid angle is shown. As the actual simulated system is two-dimensional, we only consider the upper hemisphere and then apply symmetry. Fig.~\ref{discr}-b shows the discretization of the spatial domain. The MFP distribution, shown in~\ref{discr}-c, is obtained by means of a frequency-dependent model, as described in~\cite{romano2013multiscale}.

\begin{figure}[h!]
\begin{center}
\includegraphics[width=0.7\columnwidth]{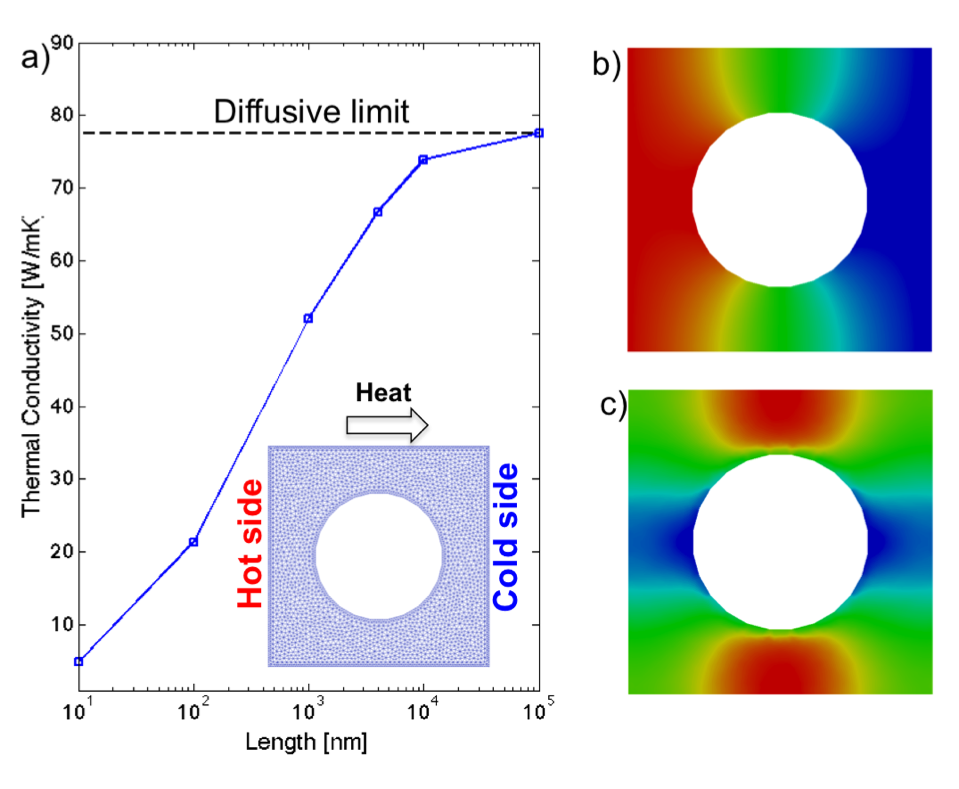}
\caption{\label{result} a) Thermal conductivity versus the length of the unit cell. For very large unit cells, thermal transport reaches the diffusive limit. In the inset, the discretization of the simulation domain is shown. Periodic boundary conditions are applied along the direction of the heat flux. b) Normalized temperature distribution c) Normalized thermal flux distribution. Due to phonon-boundary scattering, heat travels in areas that are far away from pore boundaries.}
\end{center}
\end{figure}

The phonon thermal conductivity (PTC) is computed by varying the length of the unit cell from the nanoscale to the macroscale. As shown in Fig.~\ref{result}, for very small unit cells, the thermal conductivity is reduced by roughly seven times with respect to the diffusive value, which is given by the approximated formula~\cite{Nan_1997}
\begin{equation}
\kappa_{eff} = \kappa_{bulk}\frac{1-\phi}{1+\phi},
\end{equation}
where $\phi$ is the material porosity. In our case, we choose $\phi=0.25$, which leads to the PTC $\kappa_{eff}\approx 77 W/mK$. For very large unit cells, the PTC correctly recovers the diffusive limit. In our calculations, the pore walls diffusively scatter phonons. In general, depending on the surface roughness, phonon wavelength and temperature, there might be a fraction of phonons specularly reflected~\cite{Ziman_2001}. The BTE-MFP has also been applied to the realistic case described in~\cite{Song_2004}, finding good agreement with experiments. These calculations represent a validation of the developed code and a starting point for PTC minimization in porous materials with different pore configurations. For example, in~\cite{Romano_2014} it has been shown that triangular pores arranged in misaligned columns bring a reduction in the PTC of about $60\%$ with respect to the aligned case with the same porosity.

\section{Monte Carlo methods}

One of the major limitations of traditional deterministic solution methods lies in the discretization of the 6 dimensions {\textendash} 7 dimensions for transient problems {\textendash} of the phase space. Particle Monte Carlo methods have gained popularity thanks, in part, to their ability to avoid the full discretization of the phase space. They are commonly used for solving the Boltzmann equation for a range of physics problems such as neutron transport, radiation, rarefied gases or electron transport. The idea of using such an approach for solving phonon transport problems was first introduced by Klitsner~\cite{Klitsner_1988} and further improved by Peterson~\cite{Peterson_1994} and Mazumder \textit{et al}~\cite{Mazumder_2001}. Various improvements were added between the years 2000 and 2014 and made it a powerful method.

Monte Carlo methods for particle transport solve the Boltzmann equation by {\it directly simulating} the physical processes involved.  Phonons are characterized by their position, their frequency, their polarization and their traveling direction, although alternative descriptions, using for instance the wave vectors instead of the frequencies, are also adequate. The group velocity vector is deduced from these properties, using the dispersion relation. Simulated phonon properties are randomly drawn from specified distributions in the phase space, and the rules for their evolution in time are deduced from the physical formulation. Quantities of interest, such as temperature or heat flux, are calculated by ensemble averaging. The Monte Carlo algorithm that solves for the transient non-linear BTE with the relaxation time approximation starts with an initialization step where the initial properties of a population of $N$ computational phonons are drawn from a known distribution defined by the initial condition. The evolution in time of this population is then calculated through a split algorithm where advection and scattering of phonons are treated separately. The algorithm is described in detail in~\cite{Mazumder_2001}; further details are provided in~\cite{Lacroix_2005,Hao_2009}. A timestep $\Delta t$ is defined and, starting from the positions at time $t$, positions at time $t+\Delta t$ are deduced assuming that particle trajectories are collision-free. This is the advection step. The scattering step consists of simulating the scattering events that occur between times $t$ and $t+\Delta t$. Given the relaxation time $\tau$ of a computational particle, the probability of occurrence of a scattering event may be calculated by $1-\exp(-\Delta t/\tau)$. Thus this probability law is used to randomly select particles undergoing a scattering event. Selected particles are replaced by new particles whose properties are drawn from the local “post-scattering” distribution $I^0(T)/(v \tau)$. 

The scattering process must conserve energy. In early work~\cite{Mazumder_2001,Lacroix_2005,Hao_2009}, this was traditionally done approximately by adding and deleting particles until the post-scattering energy matched the pre-scattering one. Recently, a method for enabling exact energy conservation was proposed~\cite{P_raud_2011}. It relies on the simulation of computational particles representing a fixed amount of energy instead of a fixed number of phonons. As a result, energy is rigorously conserved by simply conserving the number of computational particles. Finally, the scattering step requires the knowledge of the local temperature at every timestep. This is usually done by defining a spatial grid of computational cells in order to sample the energy density of the particle population and relating it to the temperature of the corresponding equilibrium (Bose-Einstein) distribution. This algorithm is very similar to the Direct Simulation Monte Carlo (DSMC) method used for rarefied gases~\cite{bird1994}.

The main source of error from Monte Carlo methods is the statistical uncertainty, or noise, inversely proportional to the square root of the number of independent samples (the particles) used. Noise is an important issue for problems featuring low deviations from the Bose-Einstein equilibrium since it tends to obscure the signal. Such problems are ubiquitous in nano-engineering, and yet cannot be treated by the particle Monte Carlo method presented above without resorting to massively parallel computing. Recently, a variance reduction scheme, also called “deviational” algorithm, was proposed for rarefied gases~\cite{Homolle_2007,Radtke_2012} and for phonon transport~\cite{P_raud_2011} and was shown to capture arbitrarily low deviations from equilibrium at constant computational cost. The method relies on the concept of control variates. Instead of simulating random particles representing the absolute phonon distributions, the variance-reduced algorithm simulates particles representing the {\it deviation} from a known equilibrium. In other words, if the temperature of the system is expected to feature low deviations from a given temperature $T_\text{eq}$, we can introduce the following algebraic decomposition
\begin{equation}
I = I^0(T_\text{eq}) +\left(I-I^0(T_\text{eq}) \right).
\end{equation}
to obtain the deviational BTE
\begin{equation}
\frac{1}{v} \frac{\partial I^\text{d}}{\partial t}+\mathbf{s} \cdot \nabla_\mathbf{x} I^\text{d} = \frac{I^0(T_{loc})-I^0(T_\text{eq})-I^\text{d}}{v \tau}.
\end{equation}
The deviational algorithm, which solves for $I^\text{d}=I-I^0(T_\text{eq})$ retains the main features of the non-variance-reduced algorithm and adapts them for solving the deviational BTE. Notably, $I^\text{d}$ may be negative and simulated particles must carry a sign. The particles representing the initial condition are drawn from the deviational initial distribution and the advection step of the split algorithm is unchanged. Although the selection rule for scattered particles is also unmodified, the post-scattering properties should now be drawn from the local deviational distribution $(I^0(T_\text{loc})-I^0(T_\text{eq}))/\tau$. The total energy density is found by adding the stochastic deviational energy density to the deterministically-calculated energy density corresponding the equilibrium distribution.

The variance-reduced algorithm, discussed in detail in~\cite{P_raud_2011}, offers the three following advantages:
\begin{itemize}
\item[-]	It does not introduce any approximation with respect to the traditional algorithm
\item[-]	The reduction of the statistical uncertainty, essentially results from the fact that the amplitude of the simulated deviational distribution is small with respect to $I^0(T_\text{eq})$. In particular, arbitrarily low deviations from equilibrium can be simulated with no additional cost.
\item[-]	The algebraic decomposition is inherently multiscale. It focuses the calculation effort to the regions where deviation from equilibrium is non-zero. By extending this idea to simulations of deviation from a Fourier solution, one can achieve algorithms which use particles {\it only} when the deviation from the Fourier solution is non-zero, that is where size effects are important -- the definition of a multiscale method. The simulation of a thermoreflectance experiment presented in~\cite{P_raud_2011} is a compelling example of this effect. 
\end{itemize}

Additional computational benefits can be obtained by combining the deviational formulation with the linearization of the BTE, which amounts to linearizing the collision operator as follows
\begin{equation}
\frac{(I^0(T_\text{loc})-I^0(T_\text{eq}))}{\tau(\omega,p,T)}\approx \frac{(T_\text{loc}-T_\text{eq})}{\tau(\omega,p,T_\text{eq})} \frac{\partial I^0(T_\text{eq})}{\partial T}  
\end{equation}
In other words, the distribution from which the post-scattering properties are drawn does not depend on the local temperature since, after normalization, the temperature term $T_\text{loc}-T_\text{eq}$ cancels. We recall that the algorithms presented above require the use of a timestep $\Delta t$ because knowledge of the temperature field (which features in $I^0(T_\text{loc})$) is required for simulating the scattering of the particles. Within the linearized approximation, such information is not needed. The scattering rates can be taken at temperature $T_\text{eq}$, and the post-scattering properties are all taken from a fixed distribution. Finally, we already explained that energy conservation is ensured by conserving the number of computational particles in the energy-based formulation. These observations yield the following consequences. Particles may be simulated one by one, independently from one another (a timestep is not needed). For a given particle, the time between each scattering event may be simply calculated from an exponential distribution with survival parameter $\tau(\omega,p,T_\text{eq})$. In other words, the travelling time between each scattering event is calculated by the formula $\Delta t = -\tau(\omega,p,T_\text{eq}) \ln (R)$, where $R$ is a random number uniformly drawn in the range (0,1).
The resulting “kinetic-type” algorithm is similar to existing algorithms used for neutron or phonon transport, where particles are assumed to interact with the underlying medium only. Details on the implementation may be found in Refs.~\cite{Pe_raud_2012,P_raud_2012}. Reference~\cite{Hadjiconstantinou_2014} shows that, at an equilibrium temperature of 300 K, the linearized approximation is reasonably accurate up to a deviation of 30 K.

\begin{figure}[h!]
\begin{center}
\includegraphics[width=0.7\columnwidth]{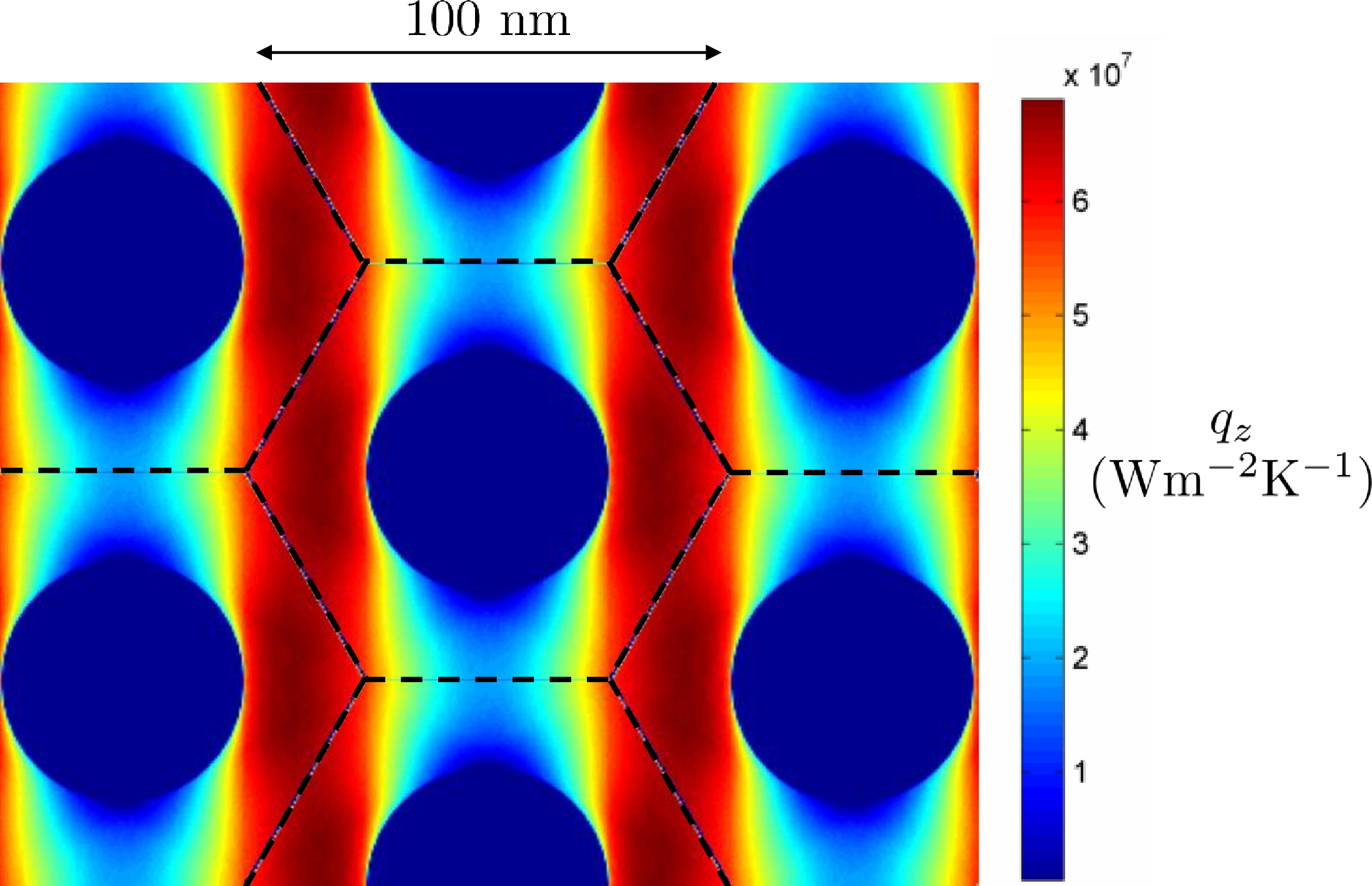}
\caption{\label{fig:periodicMC}. A component of the heat flux in a periodic nanostructure with circular pores arranged into a honeycomb structure. The materials parameters used for this simulation are the silicon parameters used in ~\cite{P_raud_2011}. The diameter of the circular pores is 50 nm. The heat flux results from an externally applied temperature gradient of $1 \times 10^6$ Km$^{-1}$ (see Ref.~\cite{Pe_raud_2012} for detailed information on the formulation that enables the simulation of an externally applied temperature gradient with periodic boundary conditions). The component of the heat flux shown here is parallel to the applied gradient. In this calculation, the boundaries of the nanopores diffusely reflect the phonons.}
\end{center}
\end{figure}

Although particles methods are inherently explicit in time, it is shown in~\cite{Pe_raud_2012} that most steady problems can be efficiently (and rigorously) treated using this approach. Traditional time-based Monte Carlo algorithms for phonon transport typically solve steady state problems by letting a time-dependent system evolve towards steady-state. Since quantities of interest are then only sampled when the steady state is reached, the number of timesteps needed to compute the transition from the initial condition to the steady state wastes significant computational resources. To the contrary, by treating each particle independently, the ``kinetic-type'' approach can deliberately ignore the transitory regime and only simulate particles at the steady state. Such particles are emitted from the steady sources only, and particle trajectories are terminated only when they leave the spatial domain, independently of how long the particle has been staying in the system. The resulting method rigorously solves the steady Boltzmann equation. Figure \ref{fig:periodicMC} shows an example of steady state calculation in a nanoporous material.

\begin{figure}[h!]
\begin{center}
\includegraphics[width=0.7\columnwidth]{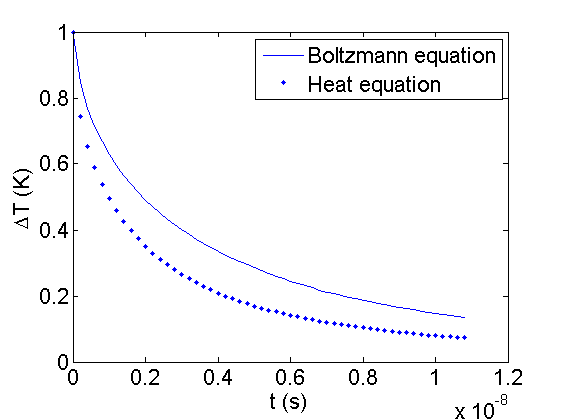}
\caption{\label{fig:transientMC}. In this figure, a homogeneous infinite material is considered (the material properties of silicon were used for the calculation). The computational domain is $\mathbf{R}^3$. The initial condition is as follows: $T(x,y,z,t=0)=301$ K within the cube defined by $0 < x < L$, $0 < y < L$ and $0 < z < L$, with $L=2$ microns, and $T(x,y,z,t=0)=300$ K outside of the cubic region.  Following this initial heating, the linearized Monte Carlo technique can be used to calculate the average temperature within the 2 microns cube against time. The deviational distribution is defined with respect to the equilibrium at temperature $T_\text{eq}=300K$. The quantity $\Delta T$ therefore corresponds to $\Delta T=T-T_\text{eq}$. In this configuration, the solution of the heat equation can be calculated analytically. The analytical solution requires the knowledge of the diffusivity, which can be computed from the materials parameters.}
\end{center}
\end{figure}

The kinetic Monte Carlo method yields two immediate advantages. It features substantial memory savings since particles are simulated one by one, and eliminating the need of a time step significantly reduces the cost of calculating each trajectory. Speedups of a factor 100 to 1000 with respect to the timestep-based deviational algorithm have been reported~\cite{Pe_raud_2012}. Finally, the method is inherently multiscale in time. Relaxation times in typical materials such as silicon are known to span several orders of magnitude. Since the time between each scattering process is computed independently of other phonons, the process automatically adapts itself to the time scale of the relaxation time of each step. It should also be noticed that this algorithm completely removes the need for spatial and time discretization. Not only does it remove the associated errors, it also allows to simulate systems of infinite sizes. Figure \ref{fig:transientMC}, which shows the time-dependent Boltzmann solution in an infinite system, highlights these features.

An important aspect of particle Monte Carlo methods lies in the fact that they primarily return {\it moments} of the underlying solution. For instance, the temperature at a given point in space is calculated in an average sense, within a volume surrounding this point. The accuracy of the estimate thus depends on the number of particle trajectories intersecting the volume and contributing to the estimate. Consequently, estimates calculated within small volumes tend to feature high statistical uncertainties. Recently, the adjoint Boltzmann transport equation for phonons was introduced as a means of addressing this limitation~\cite{Hadjiconstantinou_2014}. The adjoint BTE describes a particle problem where particles travel backward in time and where sources and detectors are switched. Thanks to this formulation, estimates can be produced in arbitrarily small volumes in the phase space, including surfaces and points. This is useful for instance for producing the contributions of individual phonon modes to the heat flux~\cite{Peraud_adjoint,Minnich2014}.

The techniques presented above focused exclusively on the relaxation time approximation which, while relatively convenient and reasonably accurate for a number of materials, fails to capture the specific features of three-phonon scattering. The three-phonon scattering operator, which incorporates both energy and momentum conservation, has been widely acknowledged to be a more accurate description~\cite{Ziman1960}. In addition, while the relaxation time approximation is semi-empirical and requires the fitting of parameters, the three-phonon scattering may be entirely derived from ab-initio calculations. Unfortunately, the resulting BTE is so complicated that very few general methods incorporating three-phonon scattering have so far been developed. Notably, efficient solution techniques for solving the {\it homogeneous} equation were recently developed ~\cite{Mingo2014}. Monte Carlo methods are well-known for their ability to solve highly complex partial differential equations and, as such, hold significant potential. Landon and Hadjiconstantinou recently developed a promising solution method based on the deviational approach for treating 2D materials such as graphene~\cite{Landon2014c}.

\section{Need for multiscale modeling}

\begin{figure}[htbp]
\begin{center}
\includegraphics[width=0.7\columnwidth]{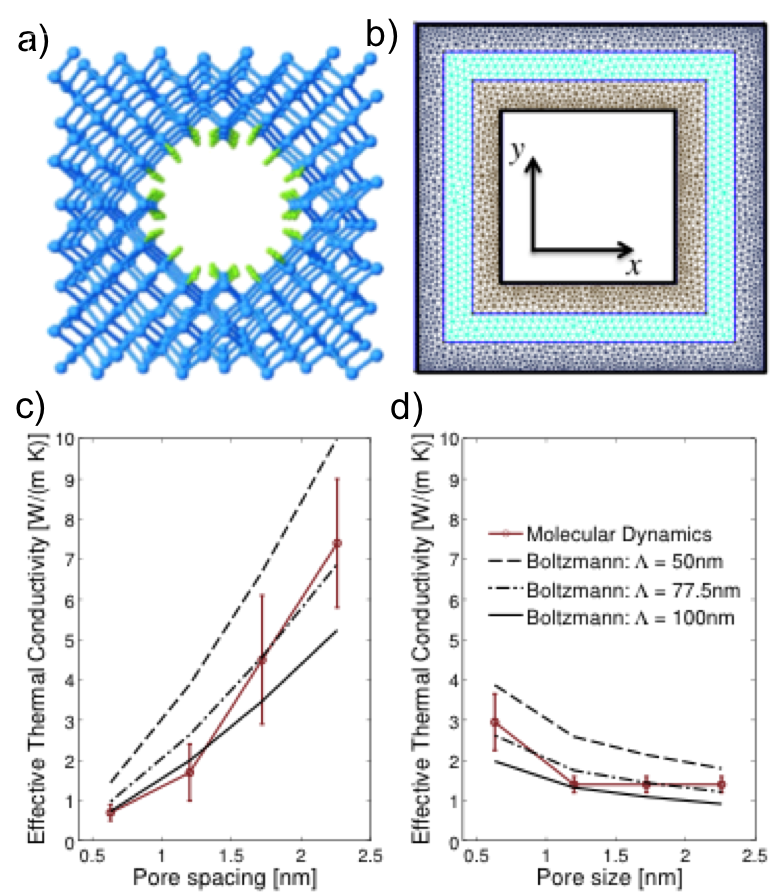}
\caption{\label{multiscale} a) Porous Silicon modeled atomistically. The pore walls have been hydrogenated. b) Porous Silicon modeled by means of a continuum domain. c) Effective thermal conductivity versus pore spacing d) Effective thermal conductivity versus pore size.}
\end{center}
\end{figure}

In the previous sections, we have described two solution techniques for BTE for nano- and mesoscale thermal transport. Also, in the very beginning we have briefly described the Molecular Dynamics method and its importance in certain nanoscale systems such as those dominated by interfaces. In Fig.~\ref{multiscale} we report a comparison between the MD and BTE calculations, varying pore distances and sizes in porous Silicon. 

Specifically for this case, we used the so called "gray approximation", where all phonons are assumed to the have the same MFP, which is used as a parameter to fit MD data. Although MD and BTE results agree qualitatively with each other, the BTE is not as accurate as MD, especially because the phonon dispersion differs from the bulk on very small scales. It is, therefore,  paramount for future research to identify the range of validity of both models, and ultimately bridge them in accurate yet computationally affordable simulations. In reference to the porous system, when pores are filled with a guest material, interfacial effects may become important and a detailed treatment of phonon transmission across such an interface is necessary. On the other hand, performing MD over the whole domain is computationally prohibitive. As a consequence, a good approach could be to split the simulation domain in two parts, one where MD is performed and one governed by the BTE. A good multiscale method should then ensure flux conservation between the two regions and define appropriate rules for the atomistic-to-continuum coupling~\cite{wagner2008atomistic}.

\clearpage

\bibliography{biblio}

\end{document}